\documentclass[reprint,twocolumn,secnumarabic,amssymb, nobibnotes, aps, pra,superscriptaddress]{revtex4-1}
\usepackage{extarrows}
\usepackage{amsmath}    % need for subequations
\usepackage{graphicx}    % need for figures
\usepackage{verbatim}    % useful for program listings
\usepackage{color}           % use if color is used in text
\usepackage{subfigure}   % use for side-by-side figures
\usepackage{hyperref}    % use for hypertext links, including those to external documents 
\usepackage{verbatim}  % and URLs
\usepackage{textcase}

\begin{document}
\title{
%Structurally-triggered 
Metal-Insulator Transition in CaCu$_3$Fe$_4$O$_{12}$ }

\author{Atsushi Hariki}
\affiliation{Department of Physics and Electronics, Graduate School of Engineering, Osaka Metropolitan University 1-1 Gakuen-cho, Nakaku, Sakai, Osaka 599-8531, Japan.}
%\affiliation{Institute of Solid State Physics, TU Wien, 1040 Vienna, Austria}
\author{Tatsuya Yamaguchi}
\affiliation{Department of Physics and Electronics, Graduate School of Engineering, Osaka Metropolitan University 1-1 Gakuen-cho, Nakaku, Sakai, Osaka 599-8531, Japan.}
\author{Mathias Winder}
\affiliation{Institute of Solid State Physics, TU Wien, 1040 Vienna, Austria}
\author{Jan Kune\v{s}}
\affiliation{Institute of Solid State Physics, TU Wien, 1040 Vienna, Austria}
%\affiliation{Institute of Physics, Czech Academy of Sciences, Na Slovance 2, 182 21 Praha 8, Czechia}
\affiliation{Department of Condensed Matter Physics, Faculty of Science, Masaryk University, Kotl\'a\v{r}sk\'a 2, 611 37 Brno,
  Czechia}
\date{\today}

\begin{abstract}
      We study structurally-triggered metal-insulator transition in CaCu$_3$Fe$_4$O$_{12}$ by means of local density approximation (LDA) +$U$ and LDA+dynamical mean-field theory (DMFT). The ferrimagnetic insulating phase is essentially the same within both approaches. While LDA+$U$ describes the metal-insulator transition as a Peierls-like instability driven
      by Fermi surface nesting in the magnetically ordered phase, LDA+DMFT allows also the site-selective Mott transition without magnetic ordering as well as smooth crossover between the two pictures. We point out similarities and differences to rare-earth nickelates.
%      We show from the LDA+$U$ calculations that the Peierls-like instability in is induced by the breathing distortion of Fe-O bonds, that is structurally triggered by rotation of FeO$_6$ octahedra. Based on the LDA+DMFT calculations, taking all Cu 3$d$, Fe 3$d$, and O 2$p$ bands into account explicitly, we reveal electronic coupling of Fe (formally tetravalent) 3$d$ and O 2$p$ states under the breathing distortion as well as origin of the ferrimagnetic ordering.
\end{abstract}

\maketitle

\section{I\lowercase{ntroduction}}
Synergy of the charge, spin, orbital or lattice degrees of freedom is known to give rise to a plethora of different thermodynamic phases and transitions between them in transition-metal oxides (TMOs)~\cite{imada98,khomskii14} with perovskite structure. Competition between crystal field, spin-orbit coupling and Hund's exchange leads to spin-state crossovers~\cite{Tanabe54,Werner07,Gavriliuk08} and
possibly to spin-state ordered phases~\cite{Kunes11,Afonso19} or excitonic magnetism~\cite{Khaliullin13,Kunes2014b,Yamaguchi17}.
Different bonding between the TM $e_g$ and $t_{2g}$ orbitals with oxygen $p$ orbitals leads to coexistence of local moments with itinerant electrons in some materials, e.g.~doped manganites~\cite{Zener51,Tokura99,Brink99}, while others are Mott insulators described by pseudospin models of Kugel-Khomskii type~\cite{Kugel82,khomskii14}. The perovskite structure built out of corner-sharing oxygen octahedra allows various structural deformations. Rotations/tilting of the oxygen octahedra~\cite{Glazer72,Woodward97,Howard98} affect the electronic bandwidths. Cooperative Jahn-Teller distortions~\cite{Jahn37,Goodenough98,khomskii14,Bersuker}
strongly couple to orbital orders~\cite{Kanamori60,Mazin07,Tokura00,Pavarini10},
while breathing distortions couple to charge order or site-selective Mott state~\cite{Park12,Greenberg18,Hariki21}.

In high-valence TMOs, which are metallic or close to metal-insulator transition (MIT), the charge degree of freedom adds to the complexity. Concomitant MIT and structural transition followed by unusual magnetic ordering attract much attention
to rare-earth nickelates RNiO$_3$~\cite{Mazin07,Lee11,Park12,Lau13,Johnston14,Jaramillo14,Subedi15,Bisogni16,Mercy17,Varignon17}. Starting from the ionic picture, Mazin~{\it et al.}~\cite{Mazin07} proposed the charge disproportionation (CD) $2d^7\rightarrow d^6+d^8$ to be the origin of a breathing distortion of NiO$_6$ octahedra and MIT. Park~{\it et al.}~\cite{Park12} used LDA+DMFT and introduced the concept
of the site-selective Mott transition. 
The CD was shown to emerge as an effective low-energy description of the site-selective Mott transition from LDA+DMFT calculations~\cite{Seth2017}. Lee~{\it et al.}~\cite{Lee11} took the band picture to argue that a Peierls-like distortion originating in Fermi surface nesting explains the physics of nickelates. This view is supported by LDA+$U$ calculations of Mercy~{\it et al.}~\cite{Mercy17} in the magnetically ordered state
who pointed out the importance of octahedral rotations for tuning the Peierls-like instability. 
In this picture, the metal-insulator transition is connected with the magnetic ordering, 
which is at odds with the paramagnet-to-paramagnet MIT transition in a number of nickelates.
The question of Peierls vs site-selective Mott mechanism~\cite{Ruppen15} remains a subject of
active debate.
%Irrespective of 
%its root cause, the structural transition was shown to be accompanied by site-selective Mott transition~\cite{Lee11,Lau13}. 
The site-selective Mott transition was reported also in Fe$_2$O$_3$~\cite{Greenberg18} under pressure.

In this Article, we study the iron oxide CaCu$_3$Fe$_4$O$_{12}$ (CCFO), which belongs to A-site ordered TM perovskites AA$_3'$B$_4$O$_{12}$~\cite{Shimakawa08,Cheng13,Long09,Meyers13,Shimakawa15,Yamada14}. 
Iron in CCFO is proposed to have anomalously high Fe$^{4+}$ valence~\cite{Yamada08}.
In addition to common parameters in TMOs, such as intra-atomic Coulomb interaction, band filling or structural distortions, the interaction between TM ions on A$'$ and B sites adds another control parameter in this group of materials.

\begin{figure}[t] 
\includegraphics[width=75mm]{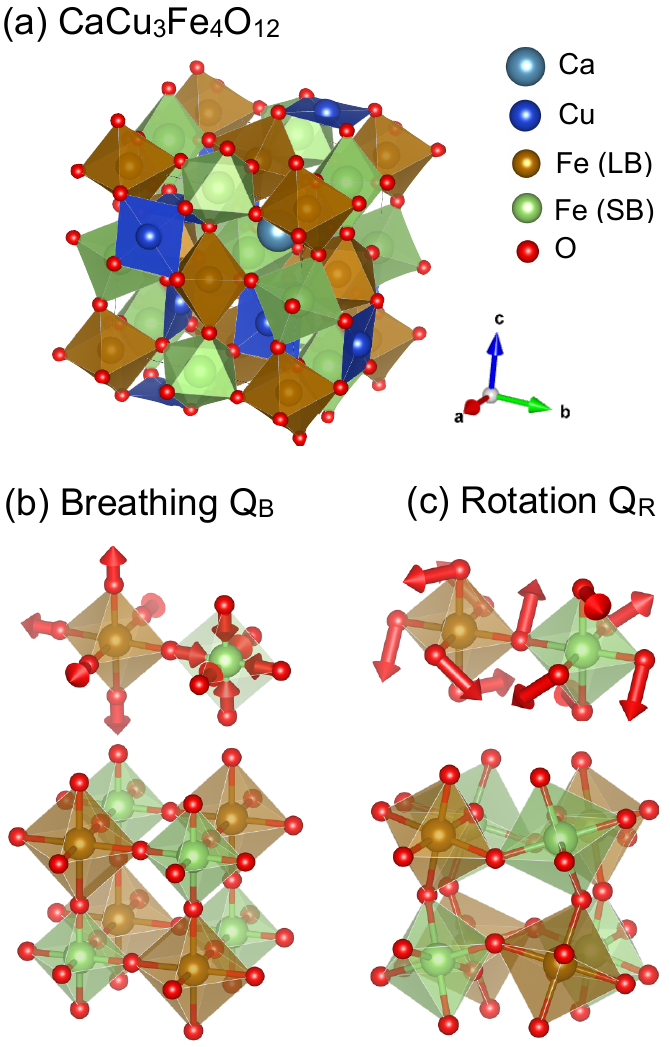}
%\caption{(a) Low-temperature structure of CaCu$_3$Fe$_4$O$_{12}$ (space group $Pn \overline{3}$). (b) breathing distortion of oxygen atoms. (c) in-phase rotation about $x$,$y$ and $z$ directions with the same amplitude ($M^{+}_{2}$). (d) elongation of the octahedra along the [111] axis ($M^{+}_{1}$). The crystal structure is visualized using VESTA3~\cite{vesta}. The mode analysis is performed using AMPLIMODE~\cite{Perez10}.
%Cu (Ca) atoms sitting on 3/4 (1/4) of the A sites in the ABO$_3$ perovskite lattice show no displacements from the high-symmetry positions in all temperatures. Thus, Co and Ca atoms are not shown in panels (b)--(d).}
\caption{(Color online) \textcolor{black}{(a) Low-temperature structure of CCFO (space group $Pn \overline{3}$). (b) Breathing distortion $Q_B$ of oxygen atoms. (c) Rotation distortion $Q_R$.}
%(d) $M^{+}_{1}$ mode. 
The crystal structure is visualized using VESTA3~\cite{vesta}. The mode analysis is performed using AMPLIMODE~\cite{Perez10}.
Cu (Ca) atoms sitting on 3/4 (1/4) of the A sites in the ABO$_3$ perovskite lattice show no displacements from the high-symmetry positions in all temperatures. Thus, Cu and Ca atoms are not shown in panels (b)--(c).}
\label{fig_struct}
\end{figure}

The high-temperature crystal structure of CCFO belongs to $Im \overline{3}$ space group with a sizable rotation of FeO$_6$ octahedra, which can be viewed as a frozen $M_{2}^{+}$ mode
%(in-phase rotation around $x$,$y$ and $z$ directions with the same amplitude,
(viewed as a rotation of the octahedra along the [111] axis) with a minor $M_{1}^{+}$ contribution (elongation of the octahedra along the [111] axis), see Supplementary Material (SM)~\cite{sm}.
The $M_{2}^{+}+M_{1}^{+}$ distortion (with the experimental amplitudes) is referred to as the $Q_{\rm R}$ distortion hereafter. 
\textcolor{black}{The $Q_{\rm R}$ distortion is shown in Fig.~\ref{fig_struct}c}.
The $Q_{\rm R}$ distortion, which leads to a square-planar coordination of Cu with the  four O neighbors (Fig.~\ref{fig_struct}a), is commonly present in the AA$_3'$B$_4$O$_{12}$ series.
At 210~K, a structural phase transition to $Pn\overline{3}$ structure takes place simultaneously with MIT and magnetic transition. Although the Fe$^{4+}$ ions may be expected to adopt the Jahn-Teller-active high-spin $t_{2g}^3e^1_{g}$ configuration, a Jahn-Teller distortion has not been reported. Instead, a breathing distortion $Q_{\rm B}$ of the FeO$_6$ octahedra was observed in the insulating phase (Fig.~\ref{fig_struct}b). 
This gives rise to two distinct Fe sites with short and long Fe--O bonds (Fig.~\ref{fig_struct}a).

The concomitant electronic transition and structural deformation in CCFO is reminiscent of MIT in RNiO$_3$ (formally Ni$^{3+}$:~$t^6_{2g}e^1_g$) accompanied by the breathing distortion of NiO$_6$ octahedra. 
Parallel to the debate on RNiO$_3$~\cite{Johnston14,Park12,Lau13,Mazin07,Mercy17,Jaramillo14,Bisogni16,Subedi15,Varignon17}, it has been suggested that a CD of the type:~2$(d^5\underline{L}) \rightarrow d^5\underline{L}^2 + d^5$ ($\underline{L}$:~oxygen hole) takes place in the insulating phase of CCFO~\cite{Yamada08,Hao09,Ueda13,Shimakawa15}.
It is worth pointing out that in nickelates the $t_{2g}$ electrons form a spin singlet ($t^6_{2g}$), while in CCFO those tend to form a stable high-spin $t^3_{2g}$ state with $S=3/2$ moment, and thus rather different magnetic behavior is to be expected.
%Interestingly, in contrast to, formally isovalent, CaFeO$_3$ exhibiting antiferromagnetic ordering~\cite{Takano77,Woodward00,Takashi00},
Indeed, along with at the MIT, a ferrimagnetic (FM) order with a large magnetization of 9.7~$\mu_{\rm B}$/f.u.~is observed~\cite{Yamada08,Mizumaki11}.
The large magnetization indicates a ferromagnetic alignment of the Fe moments antiferromagnetically coupled to the Cu$^{2+}$ ($S=1/2$) sublattice in CCFO.
%However a comprehensive picture that links the electronic, structural and magnetic properties of CaCu$_3$Fe$_4$O$_{12}$ has not been obtained so far.

We use the local-density approximation (LDA) + $U$ and LDA + dynamical mean-field theory (DMFT) approaches 
%study the interplay between  electronic and structural degrees of freedom in CCFO.
%The two methods allow us 
to investigate the electronic response to the $Q_{\rm R}$ and $Q_{\rm B}$ distortion.
%, identify the mechanism of MIT, 
We characterize the local state of Fe ions in the uniform and disproportionated phases
and determine the preferred magnetic order. 
We discuss the similarities and differences of CCFO to RNiO$_3$ in the context
of the Peierls and site-selective Mott scenarios.
%upon the bond disproportionation, and to find the magnetic order.
%We show that MIT arises from the Peierls-type instability, which triggers the breathing distortion.
%that is structurally triggered by the rotation distortion of the FeO$_6$ octahedra.
%Thus CaCu$_3$Fe$_4$O$_{12}$ has the same insulating origin in rare-earth nickelates RNiO$_3$, %and %CaFeO$_3$, proposed recently in Ref.~\cite{Mercy17} and Ref.~\cite{Zhang18}, respectively.
%proposed recently in Ref.~\cite{Mercy17}.
%\textcolor{red}{The Cu ions (Cu$^{2+}$) are electron reservoir to low-energy Fe $d$ states and determine the dominant magnetic interaction of Fe moments as the double-exchange type.}

%%%%%%%%%%%%%%%%%%%%%%%%%%%%%%%%%%%%%%%%%%%%%%%%%%%%%%%%%%%%%
%%%%%%%%%%%%%%%%%%%%%%%%%%%%%%%%%%%%%%%%%%%%%%%%%%%%%%%%%%%%%%%

\section{C\lowercase{omputational} M\lowercase{ethod}}
The LDA+$U$ calculations are performed using the augmented plane wave and local orbital (APW+lo) method implemented in the Wien2k package~\cite{wien2k}.
%by comparing to a simulation with the generalized gradient approximation (GGA).
Following previous GGA+$U$ studies of CCFO~\cite{Hao09,Ueda13}, the effective $U$ values ($U_{\rm eff}=U-J$) are chosen as $U_{\rm eff}=$~4.0~eV and 7.0~eV for Fe and Cu ions, respectively.
We have checked the robustness of the present result with respect to the choice of the exchange-correlation potential and $U_{\rm eff}$ parameters, see SM~\cite{sm}.
The reciprocal space cut-off $K_{\rm max}$ for the basis functions was chosen such that $R_{\rm MT} K_{\rm max}=7.0$. The Brillouin zone was sampled with an $8\times 8 \times 8$ mesh.
\textcolor{black}{The LDA+$U$ calculations are performed for the experimental lattice constant ($a=7.26122$~\AA) in the low-temperature insulating structure~\cite{Yamada08}. We compute the LDA+$U$ electronic bands and total energies with varying the amplitudes of the $Q_R$ and $Q_B$ distortions from the experimental values, while the lattice constant is fixed. The experimental $Q_R$ and $Q_B$ values, denoted as $100$~\% hereafter, are extracted using AMPLIMODE package~\cite{Perez10}, see SM~\cite{sm}.}

To perform LDA+DMFT calculations~\cite{kotliar06,kunes09b,georges96}, a tight-binding model spanning Cu 3$d$, Fe 3$d$ and O 2$p$ bands, constructed using the wien2wannier~\cite{wien2wannier} and wannier90~\cite{wannier90} codes, is augmented with the local Coulomb interaction on the Fe and Cu sites.
We use the interaction parameters (in eV) $(U,J)=(7.50, 0.98)$ and $(6.80, 0.80)$ for the Cu and Fe 3$d$ shells, respectively~\cite{anisimov91,Kunes09}.
To correct for the double counting of the Coulomb interaction, we shift the site energies of the Fe and Cu orbitals by the respective shell-averaged high-frequency limit (Hartree part) of the self-energy~\cite{Kunes09}. We show that our conclusions hold for a wide range of the double-counting corrections (treated as adjustable parameters), see SM~\cite{sm}.
In the DMFT self-consistent calculation, the continuous-time quantum Monte Carlo solver with density-density approximation to the on-site interaction was used to solve the auxiliary Anderson impurity model~\cite{werner06,boehnke11,hafermann12,hariki15}.
\textcolor{black}{The LDA+DMFT calculations are performed for the experimental low-temperature structure with the Fe--O bond disproportionation ($Pn\overline{3}$, $a=7.26122$~\AA) and high-temperature metallic structure with no bond disproportionation ($Im\overline{3}$, $a=7.29546$~\AA)~\cite{Yamada08}.}
%After self-consistency is achieved, physical quantities, such as local spin/charge susceptibility and local reduced density matrix, are computed by the Anderson impurity model with converged DMFT hybridization function~\cite{hariki17b,Kunes09,Ylvisaker09,Krapek12}.

%%%%%%%%%%%%%%%%%%%%%%%%%%%%%%%%%%%%%%%%%%%%%%%%%%%%%%%%%%%%%
\begin{figure}[t] 
\includegraphics[width=0.98\columnwidth]{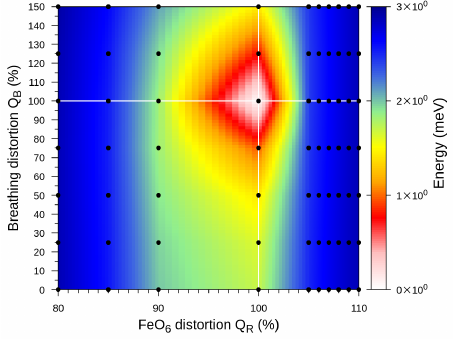}
\caption{(Color online) Evolution of the LDA+$U$ total energy in the ferrimagnetic state with the breathing distortion $Q_{\rm B}$ and $Q_{\rm R}$ rotation mode. The energies are plotted in the logarithmic scale, see SM~\cite{sm} for the energy cut for selected $Q_R$ amplitudes. Note that 100\% corresponds to the experimental amplitude.}
\label{fig_ldau1}
\end{figure}
%%%%%%%%%%%%%%%%%%%%%%%%%%%%%%%%%%%%%%%%%%%%%%%%%%%%%%%%%%%%%%%

\begin{comment}

%%%%%%%%%%%%%%%%%%%%%%%%%%%%%%%%%%%%%%%%%%%%%%%%%%%%%%%%%%%%%
\begin{figure}[h] 
\includegraphics[width=0.96\columnwidth]{3d_ene}
\caption{\textcolor{black}{{\bf We probably drop off this figure} Evolution of the total energy in the ferrimagnetic state for different lattice constants ($a=$~7.267, 7.317, 7.367, 7.417, and 7.467~$\AA$). The energy values can be found in SM~\cite{sm}. }}
\label{fig_ldau_vol}
\end{figure}
%%%%%%%%%%%%%%%%%%%%%%%%%%%%%%%%%%%%%%%%%%%%%%%%%%%%%%%%%%%%%%%

\end{comment}

\section{R\lowercase{esults}}
\subsection{LDA+$U$ and Peierls mechanism}
%First, we present the LDA+U results. 
Since the LDA+$U$ approach cannot describe fluctuating local moments, we restrict ourselves 
%this part 
to magnetically ordered phases (at $T=0$). 
%We show later that in the saturated FM state the LDA+U quasiparticle band structure is essentially identical to the LDA+DMFT one, i.e., the correlation effects in the ordered state are limited to Hartree shifts captured by LDA+U.
Overall, the FM solution has lower energy than antiferromagnetic or ferromagnetic solutions, except for $Q_{\rm R}=0$~\% where ferromagnetic alignment of all Fe and Cu moments is favored. 
Figure~\ref{fig_ldau1} shows the LDA+$U$ total energies for various amplitudes of the breathing $Q_{\rm B}$ and rotation $Q_{\rm R}$ distortions. For small $Q_{\rm R}$ the breathing
distortion is unstable, while for larger $Q_{\rm R}$ a new local minimum appears
at a finite $Q_{\rm B}$ amplitude. 
We find the global energy minimum close to the experimental $Q_{\rm R}$ and $Q_{\rm B}$ amplitudes (i.e.~$Q_{\rm R}=Q_{\rm B}=$~100~\%) irrespective of the interaction parameters within a realistic range~\cite{sm}.

%Stable ($Q_{\rm R}=0$~\%), the electronic system becomes unstable to the $Q_{\rm B}$ distortion, i.e.~appearance of the new energy minimum at a finite $Q_{\rm B}$ amplitude, progressively with the increase of the $Q_{\rm R}$ amplitude. 
%%%%%%%%%%%%%%%%%%%%%%%%%%%%%%%%%%%%%%%%%%%%%%%%%%%%%%%%%%%%%
%\begin{figure}[t] 
%\includegraphics[width=85mm]{fig_ldau1.pdf}
%\caption{(Color online) Evolution of the energy well with the breathing distortion $Q_{\rm B}$ for fixed $Q_{\rm R}$ rotation mode obtained in (a) experimental ferrimagnetic and (b) antiferromagnetic (enforced) solution. Note for $Q_{\rm R}=0$~\%, the ferromagnetic solution gets the lowest energy. The energy origin is taken at the Fermi energy $E_{F}$ (red line). (c)-(h) electronic bands (ferrimagnetic order, majority spin) for selected $Q_{\rm R}$ and $Q_{\rm B}$ amplitudes. Here, the LDA+$U$ method is employed.}
%\label{fig_ldau1}
%\end{figure}
%%%%%%%%%%%%%%%%%%%%%%%%%%%%%%%%%%%%%%%%%%%%%%%%%%%%%%%%%%%%%%%
%%%%%%%%%%%%%%%%%%%%%%%%%%%%%%%%%%%%%%%%%%%%%%%%%%%%%%%%%%%%%
\begin{figure}[t] 
\includegraphics[width=86mm]{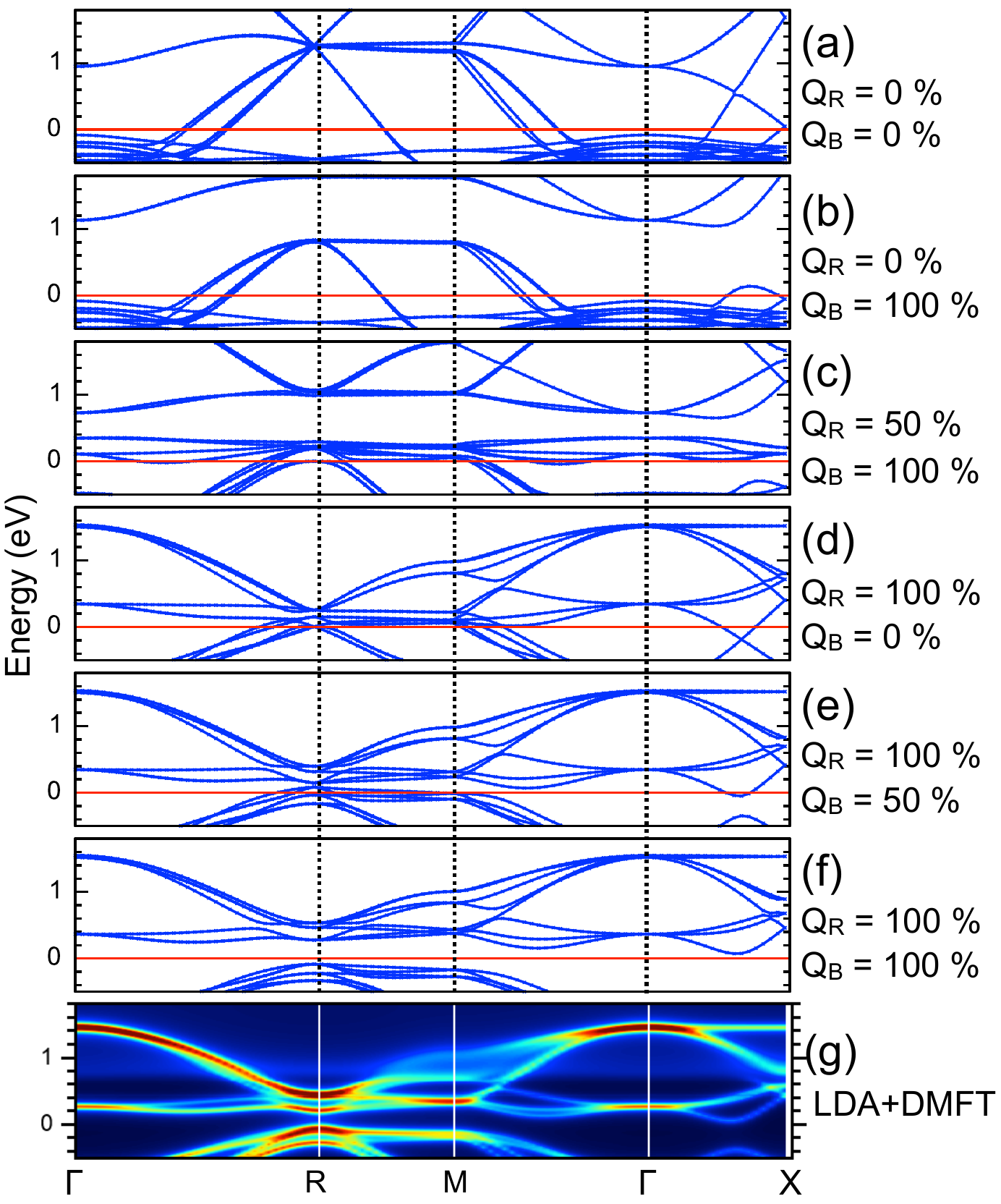}
\caption{(Color online) (a-f) Low-energy bands for the majority spin channel calculated by LDA+$U$ for selected $Q_{\rm R}$ and $Q_{\rm B}$ amplitudes. The energy origin is taken at the Fermi energy (red line). (g) LDA+DMFT spectra in the ferrimagnetic solution (majority spin). The minority-spin bands can be found in SM~\cite{sm}. }
\label{fig_band}
\end{figure}
%%%%%%%%%%%%%%%%%%%%%%%%%%%%%%%%%%%%%%%%%%%%%%%%%%%%%%%%%%%%%%%
%%%%%%%%%%%%%%%%%%%%%%%%%%%%%%%%%%%%%%%%%%%%%%%%%%%%%%%%%%%%%
\begin{figure}[t] 
\includegraphics[width=0.99\columnwidth]{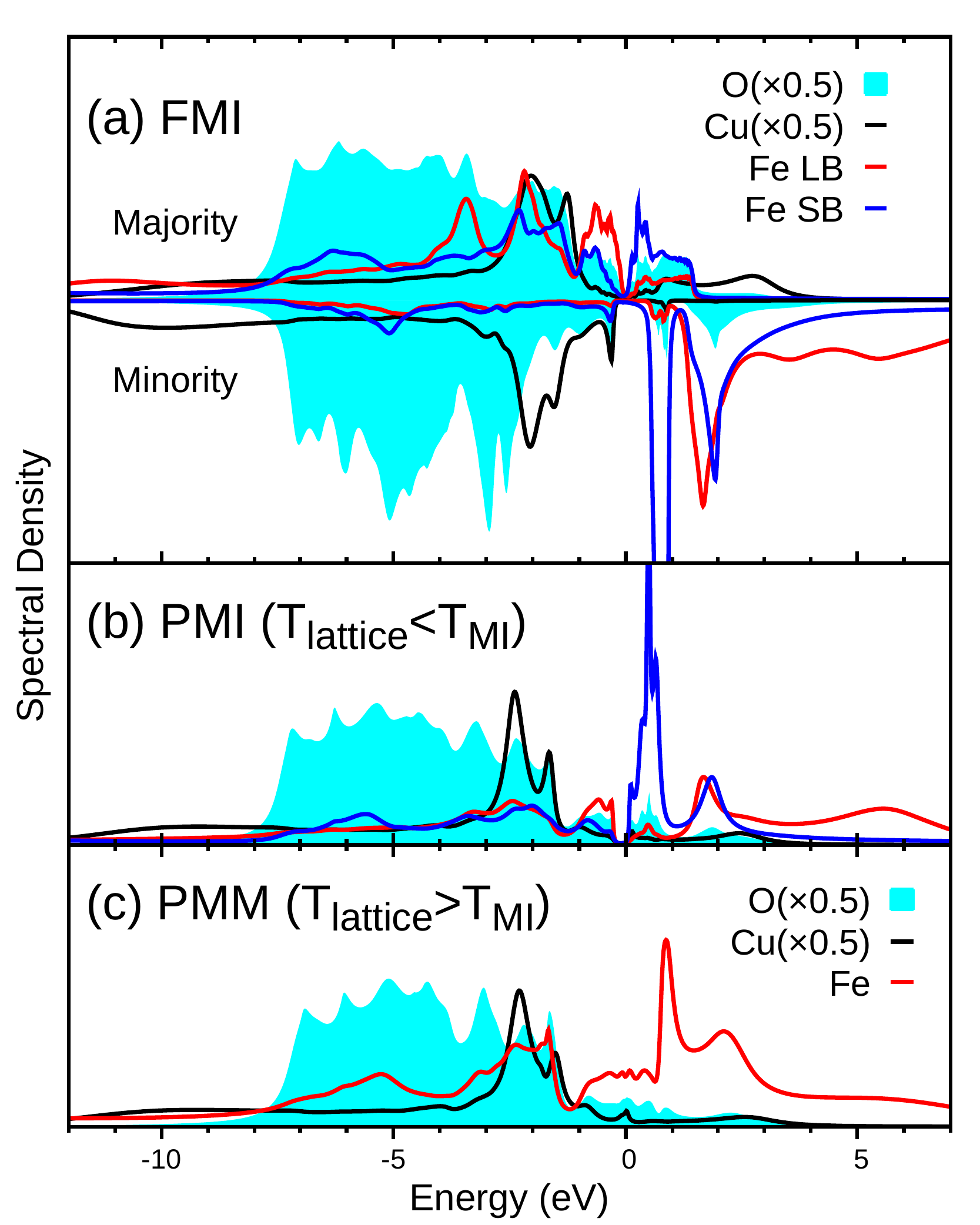}
\caption{(Color online) The $k$-integrated LDA+DMFT spectra in (a) ferrimagnetic insulating (FMI), (b) paramagnetic insulating (PMI) and (c) paramagnetic metallic (PMM) solution. The two insulating solutions are obtained with the experimental low-temperature distorted structure (with $Q_{\rm B}$), while the metallic one is obtained with the experimental high-temperature structure ($Im \overline{3}$).}
\label{fig_dos}
\end{figure}
%%%%%%%%%%%%%%%%%%%%%%%%%%%%%%%%%%%%%%%%%%%%%%%%%%%%%%%%%%%%%%%
In Figs.~\ref{fig_band}a-f we show the band structure in the vicinity of $E_F$ at selected $Q_{\rm R}$ and $Q_{\rm B}$. The depicted bands have dominantly majority-spin Fe 3$d$ $e_g$ character. The breathing distortion $Q_{\rm B}$=100~\% opens a gap for all $Q_{\rm R}$. 
However, only for $Q_{\rm R}\gtrsim$~75~\% the Fermi level falls into the gap, allowing the system to gain electronic energy and thus stabilize the breathing distortion.
The $Q_{\rm R}$ distortion reduces the $e_{\rm g}$ band width, which in turn pushes the band crossing (\@$Q_{\rm B}$=0) at $R$-point to $E_F$ (Fig.~\ref{fig_band}d).
While at $Q_{\rm R}$=100~\% the band crossing locates at the optimal position on the Fermi level, the insulating band structure is obtained already for $Q_{\rm R} \approx$~75~\%.
This picture is the Peierls-like instability for opening the gap, described in Refs.~\onlinecite{Lee11,Mercy17}, in rare-earth nickelates.

The problem with this picture is that saturated spin polarization of the Fe $e_g$ bands is necessary to open a gap at $E_F$~\cite{Mercy17}, while experimentally the normal phase is paramagnetic.
This criticism in CCFO is weaker than in case of the paramagnetic-metal to paramagnetic-insulator transition in RNiO$_3$ since short range magnetic order is expected above $T_c$ in CCFO, where MIT and magnetic ordering take place simultaneously. 

%The $Q_{\rm B}$ distortion then opens a gap at $E_F$, which is connected with a gain of electronic energy that stabilizes the new structure. 
%This substantially increases the energy gain of the electronic system by opening the $Q_{\rm B}$ gap, leading to the softening of the $Q_{\rm B}$ mode observed in Fig~\ref{fig_ldau1}.
%At the experimental $Q_{\rm R}$ amplitude, Fig.~\ref{fig_ldau1}d, the $e_{g}$ bands locates exactly on the $E_F$.
%There the electronic system becomes unstable to the $Q_{\rm B}$ mode and favors the breathing distortion $Q_{\rm B}$, that leads to the gap opening, see Fig.~\ref{fig_ldau1}ef. 

%%%%%%%%%%%%%%%%%%%%%%%%%%%%%%%%%%%%%%%%%%%%%%%%%%%%%%%%%%%%%
\begin{figure}[t] 
\includegraphics[width=0.99\columnwidth]{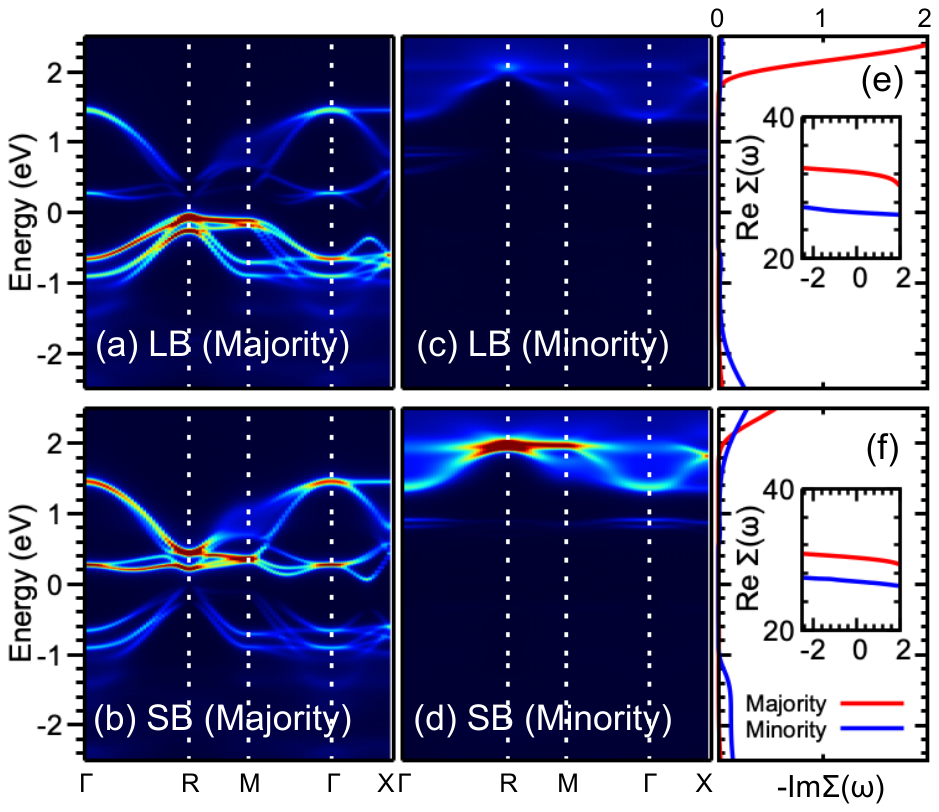}
%\caption{Low-energy valence spectra for the majority (left) and minority (right) spin in the ferrimagnetic solution obtained by LDA+DMFT. The calculation is performed for the experimental distorted structure. The energy origin is taken at the Fermi energy. The spectra projected on Fe $d$-states on long-bond (LB) and short-bond (SB) site are shown in the panels.}
\caption{(Color online) LDA+DMFT valence spectra projected onto Fe $e_{g\sigma}$-states on (a,c) long-bond (LB) and (b,d) short-bond (SB) site.
The panels (a,b) and (c,d) are for the majority-spin and minority-spin channel in the ferrimagnetic solution, respectively.
Total as well as Cu $d$-projected spectra can be found in SM.
The imaginary part of the self-energies in the real-frequency domain for (e) LB and (f) SB site. The real part of the self-energies is shown in the inset.}
\label{fig_fm1}
\end{figure}
%%%%%%%%%%%%%%%%%%%%%%%%%%%%%%%%%%%%%%%%%%%%%%%%%%%%%%%%%%%%%%%

%%%%%%%%%%%%%%%%%%%%%%%%%%%%%%%%%%%%%%%%%%%%%%%%%%%%%%%%%%%%%
\begin{figure}[t] 
\includegraphics[width=0.99\columnwidth]{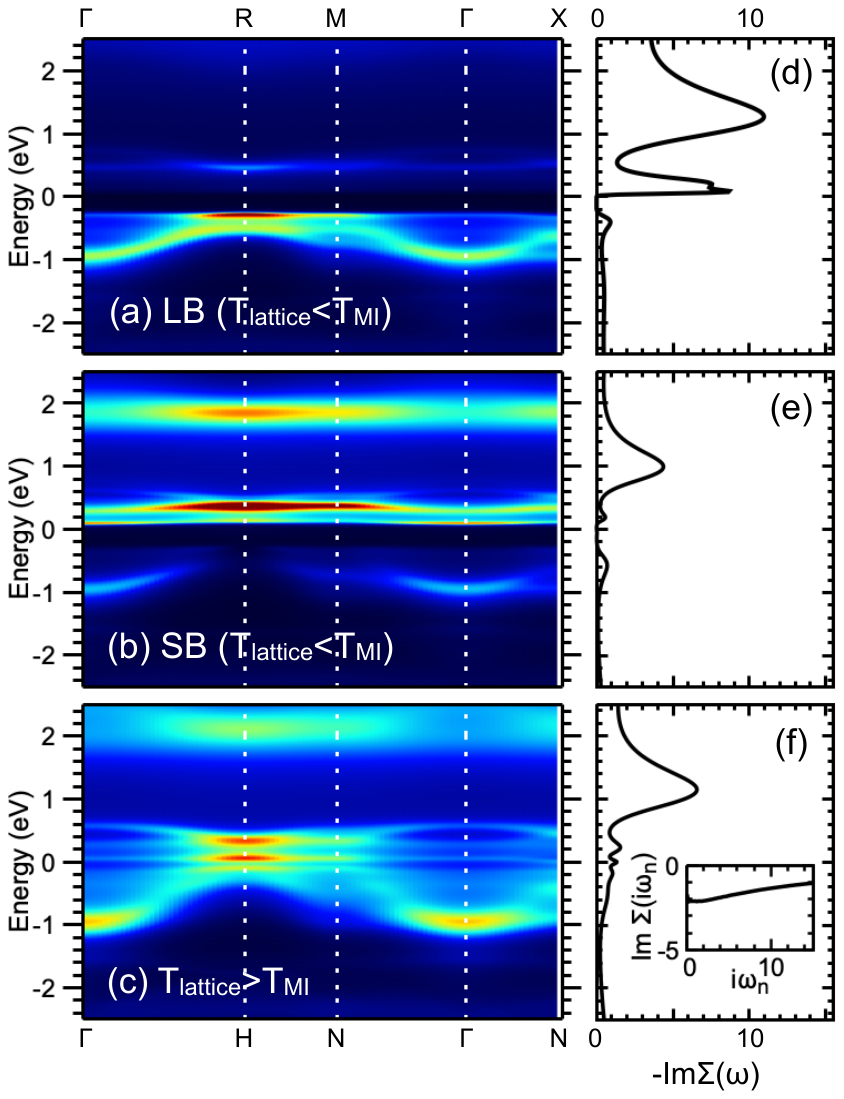}
%\caption{Low-energy valence spectra in the paramagnetic solution obtained by LDA+DMFT for the experimental distorted structure. The energy origin is taken at the Fermi energy. The spectra projected on (b) Cu and Fe $d$-states on (c) long-bond and (d) short-bond site.}
\caption{(Color online) LDA+DMFT valence spectra projected onto Fe $e_{g\sigma}$-states on (a) long-bond (LB) and (b) short-bond (SB) site in the paramagnetic insulating solution obtained with the distorted structure (i.e.~$Q_{\rm B}=Q_{\rm R}=100$~\%) (c) Fe $e_{g\sigma}$-states in the paramagnetic metallic solution obtained for the \textcolor{black}{high-temperature non-disproportionated structure} ($Im \overline{3}$).
The imaginary part of the self-energies in the real-frequency domain is shown in panels~(d-f). The inset in panel~(f) shows the self-energy in the Matsubara-frequency domain.}
\label{fig_pm}
\end{figure}
%%%%%%%%%%%%%%%%%%%%%%%%%%%%%%%%%%%%%%%%%%%%%%%%%%%%%%%%%%%%%%%
\subsection{LDA+DMFT and site-selective Mott transition}
The LDA+DMFT approach can capture the magnetically ordered phase as well as the paramagnetic state with fluctuating local moments. Our implementation of LDA+DMFT does not allow a reliable calculation of the free energy. Therefore we do not search for the most stable state as with LDA+$U$, but only investigate the electronic structure above and below the experimental structural transition. In the following we analyze the ferrimagnetic and paramagnetic solutions. We find the ferrimagnetic solution to be stable for both distorted and undistorted structure up to 1200~K, which grossly overestimates the experimental $T_c$. We attribute it largely to the density-density approximation on the on-site interaction, in which the large Fe moments are treated as Ising rather than Heisenberg moments. The PM solutions are obtained by constraint for the DMFT self-energy of the Cu and Fe $d$ electrons. 

%\textcolor{black}{
%Fig.~\ref{fig_band}g shows the low-energy spectra in the ferrimagnetic state obtained by LDA+DMFT for the experimental crystal structure ($Q_{\rm R}=Q_{\rm B}=$~100~\%) and $\mu_{\rm dc}$=58.4 (29.3)~eV for Cu (Fe) $d$-states.
%The LDA+DMFT spectra resemble the LDA+$U$ spectra of Fig.~\ref{fig_band}f with a direct gap at the $R$ point opened within the majority spin Fe 3$d$ states.

Figures~\ref{fig_dos}ab show the $k$-integrated spectra of the ferrimagnetic and paramagnetic LDA+DMFT solutions for the experimental crystal structure. In agreement with experiment, in Fig.~\ref{fig_dos}c we find a paramagnetic metallic solution in \textcolor{black}{the high-temperature non-disproportionated structure} (the FM solution not shown here is also metallic). The states around $E_F$ have dominantly Fe $e_g$ character. Unlike in RNiO$_3$, we find a rather strong damping of the $e_g$ states due to interaction with fluctuating $t_{2g}$ moments (absent in RNiO$_3$). This explains  smaller conductivity of metallic CCFO~\cite{Shimakawa15}.

The LDA+DMFT solution in \textcolor{black}{the low-temperature disproportionated structure} is insulating irrespective of the presence of the magnetic order. The FM state at the simulation temperature is fully saturated. The absence of spin fluctuations leads to a band structure that closely resembles the LDA+$U$ one, see Figs.~\ref{fig_band}fg. This is not surprising since the self-energy on both the SB and LB site, shown in Figs.~\ref{fig_fm1}ef, is dominated by the frequency-independent Hartree term with vanishing imaginary part around $E_F$. The gap then has a Peierls character between the LB-dominated valence and SB-dominated conduction band, Figs.~\ref{fig_fm1}ab, and the minority bands removed from the vicinity of $E_F$, Figs.~\ref{fig_fm1}cd.

The PM spectra in the disproportionated structure offer a rather different picture, Figs.~\ref{fig_pm}abde. The Fe $e_g$ bands are strongly damped and renormalized due to fluctuating $t_{2g}$ moments. However, the main difference is the nature of the gap that is opened due to a pole in the LB self-energy, while the SB self-energy has a Fermi liquid character. The MIT in the PM phase thus has the site-selective Mott character~\cite{Park12,Ruppen15}.

\begin{figure}[t] 
\includegraphics[width=0.98\columnwidth]{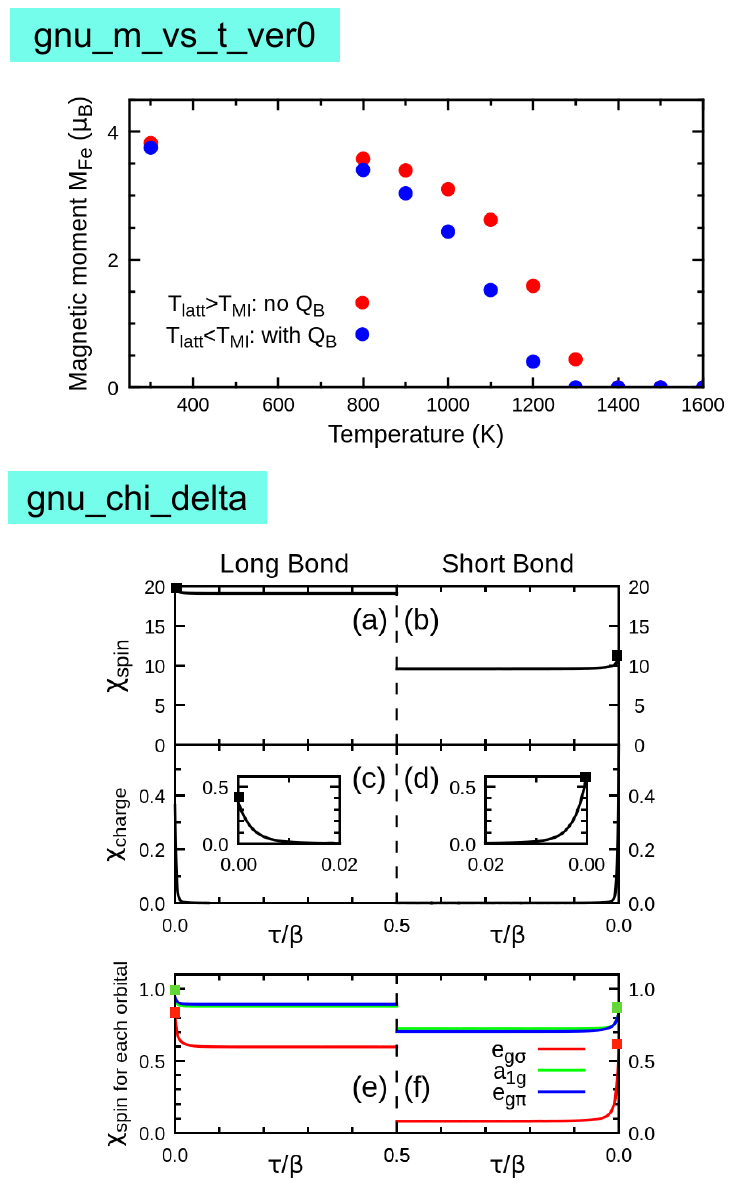}
\caption{(Color online) Local spin/charge susceptibility in (a)/(c) long-bond and (b)/(d) short-bond Fe site calculated by LDA+DMFT. The orbital-diagonal contributions in the local spin susceptibility of (e) long-bond and (f) short-bond Fe site. Note that the rotation of the FeO$_6$ octahedra splits $t_{2g}$ states into $a_{1g}$ and doubly-degenerated $e_{g\pi}$ states. Square symbols in (e) and (d) mark the value at $\tau$=0.}
\label{fig_chi}
\end{figure}
%%%%%%%%%%%%%%%%%%%%%%%%%%%%%%%%%%%%%%%%%%%%%%%%%%%%%%%%%%%%%%%

\section{D\lowercase{iscussion}}
We start our discussion by analysing the local physics on the Fe sites in the site-selective Mott state.
%Next, we discuss the local spin and charge dynamics of Fe 3$d$ electrons.
The LDA+DMFT calculation yields similar $d$--occupations (per site) of 5.18 and 5.08 on the LB and SB Fe sites, respectively.
Although the charge fluctuation is slightly larger on the SB site, the local charge susceptibilities (charge-charge correlation function) on the two Fe sites, Figs.~\ref{fig_chi}cd, resemble each other with almost incompressible Fe $3d$ states. The different nature of LB and SB sites is revealed by the local spin susceptibilities in Figs.~\ref{fig_chi}ab, which yield
the effective screened moments $m_{\rm scr}=\sqrt{T\chi_{\rm loc}}$ of 4.37 and 3.10 $\mu_{\rm B}$ for the LB and SB sites, respectively.
The long-lived moment on the SB site is in qualitative contrast with that in RNiO$_3$ with vanishing $m_{\rm scr}$ on the SB Ni site~\cite{Park12,Subedi15}.
The orbital-diagonal contributions in the local spin susceptibility
in Figs.~\ref{fig_chi}ef reveal that, while all $d$ electrons contribute to the LB local moment, only the $t_{2g}$ ones contribute on the SB site.
%show the orbital-diagonal contributions in the local spin susceptibility. All electrons on different orbitals contribute to the spin response on the long-bond Fe site, while only $t_{2g}$ electrons contribute in the short-bond Fe site.
%Thus the long-lived spin moment in the short-bond Fe site is composed of $t_{2g}$ electrons.
%The spin moment of two $e_g$ electrons, that is present on a short-time scale as exemplified by a sizable response at $\tau$=0, is  screened by coupling to oxygen via the $\sigma$-bond.
This leads to the following site-disproportionation picture of CCFO:~2$(d^5\underline{L}) \rightarrow  t^3_{2g} [e^2_g\underline{L}_{\sigma}^2] \ (S=3/2)+ t^3_{2g}e^2_g \  (S=5/2)$, where $\underline{L}_{\sigma}$ denotes a ligand hole that couples with the $e_g$ states.
%or equivalently    
%$\tilde{t}^3_{2g}\tilde{e}^0_g + \tilde{t}^3_{2g}\tilde{e}^2_g$, where $\tilde{e}_g$ ($\tilde{t}_{2g}$) represents the low-energy anti-bonding states composed of atomic $e_g$ ($t_{2g}$) and oxygen 2$p$ states.
%The CD is visible in low-energy states, see Fig.~\label{fig_fm}.

\begin{figure}[t] 
\includegraphics[width=0.96\columnwidth]{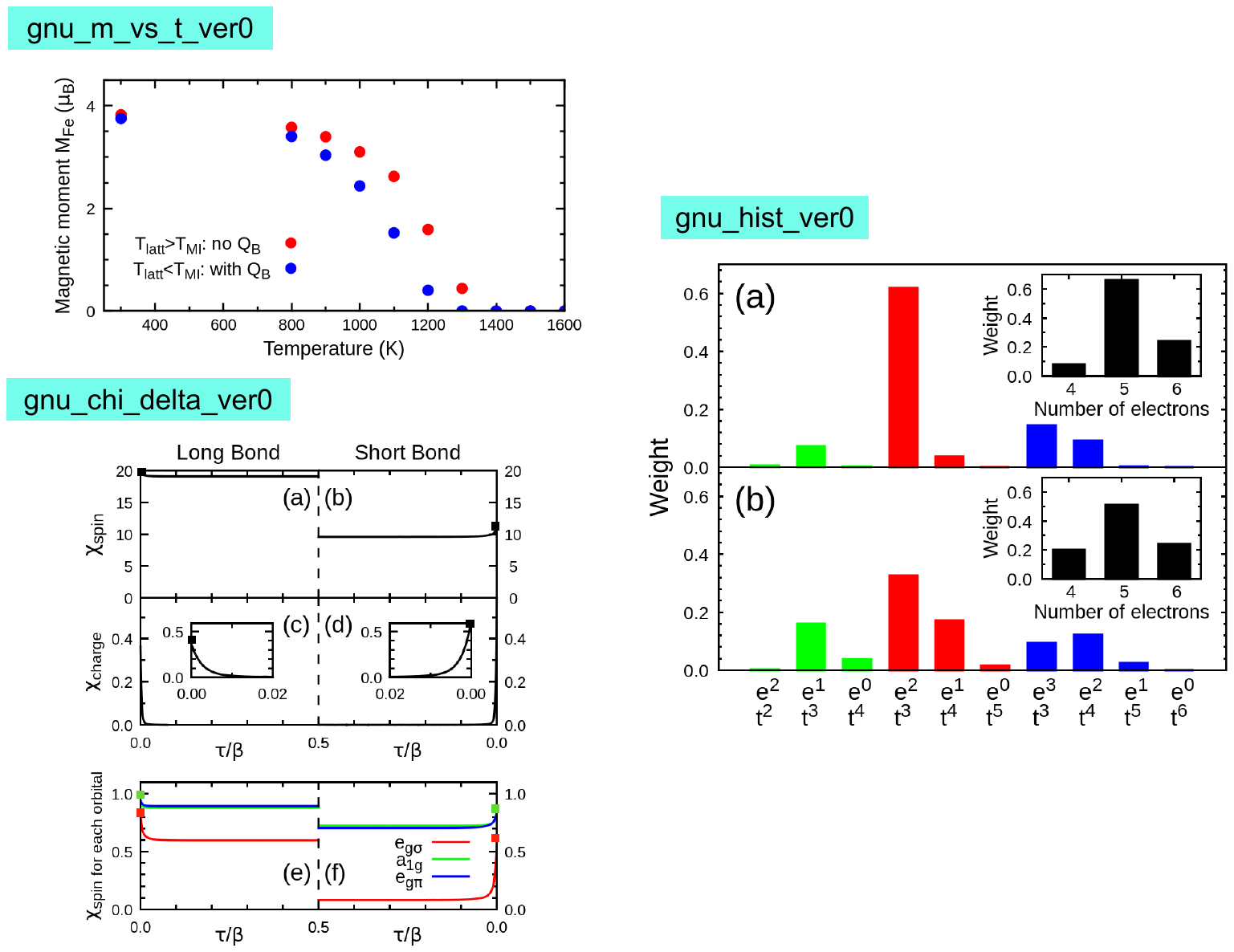}
\caption{(Color online) Weights of dominant atomic states in $N$=4 (green), 5 (red) and 6 (black) occupations on (a) long-bond and (b) short-bond Fe site. The atomic weights integrated in $N$=4, 5 and 6 sectors are shown in inset.}
\label{fig_hist}
\end{figure}

This picture is corroborated by the weights of atomic configurations
(diagonal element of the site-reduced density matrix) in 
Fig.~\ref{fig_hist}.
%shows the atomic-state weights (eigenvalues of the site-reduced density matrix) calculated for selected multiplet states in $N$=4, 5 and 6 occupations.
The high-spin ($S=5/2$) state with $e^2_g t^3_{2g}$ configuration is dominant on the LB site.
The SB Fe site shows a broader distribution of the atomic weights. This is to be attributed
to the stronger covalent bonding with O sites. While this affects mostly the $e_g$ orbitals,
the abundance of the $e^1_g t^4_{2g}$ configuration shows that reducing the role of
$t_{2g}$ electrons to building of a rigid local moment is an approximation.

%especially among states with the same $t_{2g}$ occupations but varying $e_g$ occupations, that supports the CD picture above.
%\textcolor{red}{Perhaps, the weights in the $e^1_g t^4_{2g}$ states originates from an enhanced crystal-field splitting (due to stronger covalency) in the short-bond Fe site, rather than from actual fluctuation of $t_{2g}$ electrons. Or in a similar sense, perhaps first $e^3_g t^3_{2g}$ weights are transferred to $e^2_g t^4_{2g}$ in the $N$=6 sector due to the large crystal-field. Then the following charge-fluctuation of $e_g$ electrons leads to the enhancement of $e^1_g t^4_{2g}$ in the $N$=5 sector.}

%%%%%%%%%%%%%%%%%%%%%%%%%%%%%%%%%%%%%%%%%%%%%%%%%%%%%%%%%%%%%
%%%%%%%%%%%%%%%%%%%%%%%%%%%%%%%%%%%%%%%%%%%%%%%%%%%%%%%%%%%%%%%

\begin{comment}

%%%%%%%%%%%%%%%%%%%%%%%%%%%%%%%%%%%%%%%%%%%%%%%%%%%%%%%%%%%%%
\begin{figure}[t] 
\includegraphics[width=0.99\columnwidth]{fig_tn.pdf}
\caption{\textcolor{black}{{\bf We do not mention this figure in the manuscript. Maybe we drop this off?} Magnetic moment $M_{\rm Fe}$ on Fe 3$d$ states as a function of temperature obtained by LDA+DMFT. The red (black) symbol shows the $M_{\rm Fe}$ value calculated for $Im\overline{3}$  ($T_{\rm lattice}>T_{\rm MI}$, no $Q_{\rm B}$) and $Pn\overline{3}$ structure ($T_{\rm lattice}<T_{\rm MI}$, finite $Q_{\rm B}$).}} 
\label{fig_tn}
\end{figure}
%%%%%%%%%%%%%%%%%%%%%%%%%%%%%%%%%%%%%%%%%%%%%%%%%%%%%%%%%%%%%%%

\end{comment}

Next, we discuss the origin of the magnetic ordering. We find that both in the non-disproportionated metallic state and the disproportionated insulating state
the Fe moment favor parallel (ferro- or ferrimagnetic) orientation.
To understand this, we refer to the general picture of Zener double-exchange model~\cite{Zener51} in which antiferromagneticaly
arranged local ($t_{2g}$) moments generate, via intra-atomic Hund's coupling, a staggered
potential for the electrons in partially filled $e_g$ bands. This increases the 
kinetic energy of the $e_g$ electrons and thus favors FM alignment of the local moments.
This mechanism, which
is traditionally used to describe doped manganites, survives also
in metallic systems with moderate fluctuations of $t_{2g}$ charge~\cite{Sotnikov18} such as SrCoO$_3$~\cite{Kunes12}. The metallic phase of CCFO is in this respect similar to the high-valence SrCoO$_3$.  

In the disproportionated state, the (staggered) antiferromagnetic potential leads to closing of the gap and thus insulator-compatible FM order is favored.
%The observation above suggests double-exchange-type interaction as the leading mechanism of the ferrimagnetic ordering in CCFO. The hopping of $e_g$ electrons on long-bond Fe site to short-bond on the background of the localized $S=3/2$ moment favors the ferromagnetic alignment of the Fe moments due to the local ferromagnetic Hund's coupling between the $e_g$ and $t_{2g}$ electrons.
The antiferromagnetic coupling of the Cu ($S=1/2$) and Fe moments observed in our calculations
%, most likely due to the super-exchange interaction, builds on the ferromagnetically-aligned Fe moments that surrounds the Cu moment, but the Cu moment 
is not the glue of the ferromagnetic coupling of the Fe moments.
A DMFT calculation with a paramagnetic constraint on the Cu site yields a ferrimagnetic insulating solution with almost unchanged Fe bands, see SM~\cite{sm}.

%\textcolor{red}{It would be worth noting that the simulation eliminating the Hund's coupling of Fe $e_g$ and $t_{2g}$ electrons in the local Coulomb vertices results in a ferromagnetic metallic solution, where both two Fe ions take $e^2_{g,\uparrow}t^3_{2g,\downarrow}$ configuration and Cu ion loses its spin polarization.}}

The key question concerning materials exhibiting a bond disproportionation such as RNiO$_3$ or the present CCFO is the site-selective Mott vs Peierls mechanism of the MIT. In the context of RNiO$_3$, this question was addressed in a number of recent studies~\cite{Park12,Ruppen15,Mercy17}. Due to the limitation of our present
approach we are not able to provide a definite answer - magnetic ordering and site disproportionation take place simultaneously in CCFO. Nevertheless, the strikingly different band structures of FM and PM states in Fig.~\ref{fig_fm1} and Fig.~\ref{fig_pm}
open a question whether these are, in principle, compatible with each other, i.e., can one go from the PM band structure to the FM one without closing the gap? The gap in the PM state relies on the pole in the self-energy of the LB site, while in the FM state the Hartree (energy independent) shift of the spin-minority band above the chemical potential is crucial.

To understand how the PM Mott state can evolve in the FM one without closing the gap, we invoke a toy model of Ref.~\onlinecite{Ruppen15} that captures essentials of the site-disproportionated Mott state and the behavior of the self-energy pole when going from PM to spin-polarized Mott insulator. The authors of Ref.~\onlinecite{Sangiovanni06} showed that the magnetic ordering (in one-band Hubbard model) introduces, in addition to the spin-dependent Hartree contribution to the self-energy, 
a spin dependent frequency shift of the self-energy along the frequencies. This way the LB and SB site self-energies evolve from the PM state, Figs.~\ref{fig_pm}de, to the FM state, Figs.~\ref{fig_fm1}ef. Introducing this behavior by hand to the toy model, one can go smoothly from the PM to the FM state without moving the gap away from the $E_F$ as demonstrated by a movie in SM~\cite{sm}.

%\textcolor{black}{Finally, we comment on a role of Cu ions in ACu$_3$Fe$_4$O$_{12}$ series. Though the valency of Cu in CCFO is confirmed to be divalent by e.g.~Cu $L$-edge x-ray absorption experiment~\cite{Mizumaki11}, substitution of A-site by rare-earth elements induces a dramatic electronic and structural transition from a ferrimagnetic insulator with Fe-O breathing distortion (A=Dy--Lu) to an antiferromagnetic insulator without breathing distortion (A=La, Pr--Nd, Sm--Tb), in which Cu valency is proposed to change from divalent to trivalent.
%When the Cu valency departs from divalent to trivalent, see SM~\cite{sm} (by controlling Cu $d$ level in the simulation), corresponding to effective electron doping to Fe (or to be more explicit, to anti-bonding states composed of Fe $d$ and O 2$p$ states), the ferrimagnetic-insulating solution breaks down into the metallic one in the experimental structure, though the antiferromagnetic solution is not compatible with the experimental breathing distortion in the studied parameter range.  
%Thus the Cu plays a role of charge reservoir....
%\textcolor{red}{Last try if I could get the AF insulator solution with/without rotation/breathing distortion under a Cu$^{3+}$-like setting (by using a shallower double-counting) is running and will be concluded soon.}
%}

\section{C\lowercase{onclusion}}

By means of LDA+$U$ and LDA+DMFT methods, we investigated insulating and magnetic properties in CaCu$_3$Fe$_4$O$_{12}$, consisting of high Fe$^{4+}$ valence.
By a systematic LDA+$U$ calculation combined with a distortion-mode analysis, we revealed an underlying Peierls instability to open the electronic gap in the ferrimagnetic ordered phase. The LDA+DMFT simulation indicates that the site-selective Mott mechanism also opens the insulating gap in the experimental distorted structure.
In the site-disproportionated state of CaCu$_3$Fe$_4$O$_{12}$, the two $e_g$ electrons on the short-bond Fe site couples with oxygen $p$ holes, forming a singlet, while $t_{2g}$ moments are robust in both long- and short-bond Fe sites, leading to the site-disproportionation picture:~2$(d^5\underline{L}) \rightarrow  t^3_{2g} [e^2_g\underline{L}_{\sigma}^2] \ (S=3/2)+ t^3_{2g}e^2_g \  (S=5/2)$.
In contrast to rare-earth nickelates RNiO$_3$, the paramagnetic site-selective Mott phase is not observed experimentally. The active Fe $t_{2g}$ moment ($S=3/2$), absent in RNiO$_3$, stabilizes the ferrimagnetic order via the double-exchange mechanism. 
We discussed the spectral change from the site-selective Mott to ferrimagnetic Peierls-like insulator.

%%%%%%%%%%%%%%%%%%%%%%%%%%%%%%%%%%%%%%%%%%%%%%%%%%%%%%%%%%%%%%%%%%%
\begin{acknowledgments}
We thank A. Georges, I. Yamada, M. Mizumaki, Y. Uchimura, J. Fern\'andez Afonso, A. Sotnikov, and M. Furo  for fruitful discussions. 
This work was supported by the European Research
Council (ERC) under the European Union’s Horizon 2020 research and
innovation programme, Grant Agreement No. 646807-EXMAG, (A.H., M.W., J.K.),
QUAST-FOR5249 project I 5868-N (J.K.) of
the Austrian Science Fund (FWF) and JSPS KAKENHI with Grant Numbers 23K03324 and 23H03817 (A.H.).
The computational calculations were performed at the Vienna Scientific Cluster (VSC).
\end{acknowledgments}
%%%%%%%%%%%%%%%%%%%%%%%%%%%%%%%%%%%%%%%%%%%%%%%%%%%%%%%%%%%%%%%%%%%

\bibliography{ccfo}

\end{document}